\documentclass[3p,times]{elsarticle}

\usepackage{ecrc}

\usepackage{epstopdf}

\volume{00}

\firstpage{1}

\journalname{Nuclear Physics A}

\runauth{Xiaofeng Luo}

\jid{nupha}

\jnltitlelogo{Nuclear Physics A}

\usepackage{graphicx}
\usepackage{amsmath,amssymb}
\usepackage{multirow}

\newcommand{ \pt }{\mbox{$p_{T}$}}

\newcommand{\sNN}{{{$\sqrt{s_{_{{NN}}}}$}}}

\newcommand{\KV}{{\mbox{$\kappa\sigma^{2}$}}}
\newcommand{\SD}{{\mbox{$S\sigma$}}}

\begin{document}

\begin{frontmatter}



\title{Exploring the QCD Phase Structure with Beam Energy Scan in Heavy-ion Collisions }


\author[]{Xiaofeng Luo }
\ead{xfluo@mail.ccnu.edu.cn}
\address{Institute of Particle Physics and Key Laboratory of Quark \& Lepton Physics (MOE), \\Central China Normal University, Wuhan, 430079, China. }

\begin{abstract}
Beam energy scan programs in heavy-ion collisions aim to explore the QCD phase structure at high baryon density. Sensitive observables are applied to probe 
the signatures of the QCD phase transition and critical point in heavy-ion collisions at RHIC and SPS.  Intriguing structures, such as dip, peak and oscillation,  have been observed in the energy dependence of various observables. In this paper, an overview is given and corresponding physics implications will be discussed for the experimental highlights from the beam energy scan programs at the STAR, PHENIX and NA61/SHINE experiments. Furthermore, the beam energy scan phase II at RHIC (2019-2020) and other future experimental facilities for studying the physics at low energies will be also discussed.

\end{abstract}
 
\begin{keyword}
QCD Phase Structure \sep QCD Critical Point \sep Beam Energy Scan \sep Heavy-ion Collision \sep QCD Phase Transition

\end{keyword}

\end{frontmatter}



\section{Introduction}
\label{intro}
One of the main goals of heavy-ion collisions is to explore the phase structure of hot and dense nuclear matters~\cite{bes}, which can be displayed in the phase diagram of Quantum Chromodynamics (QCD-theory of strong interactions).
Figure 1 shows the conjectured QCD phase diagram on the basis of the theoretical calculations. It can be expressed into a two dimensional diagram, the baryon chemical potential ($\mu_B$, $X$-axis) and temperature ($T$, $Y$-axis). During the past two decades, many promising signatures in high energy heavy-ion collisions have been observed for the formation of strongly interacting matter with partonic degree of freedom~\cite{2005_STARwhitepaper,2005_PHENIXwhitepaper} - the so called Quark Gluon Plasma (QGP).  The ab-initio Lattice QCD calculations 
show that at small baryon chemical potential and high temperature,  the transition from the hadron gas phase to the QGP phase is a smooth crossover~\cite{crossover}, whereas a first order phase transition is expected at high baryon chemical region~\cite{firstorder}. The endpoint of the first order phase boundary towards the crossover region is so called the QCD critical point~\cite{QCP_Prediction}. There are large uncertainties from theoretical calculations in determining the QCD phase structure at high baryon density region. For e.g,  the location of the critical point in the QCD phase diagram. Finding the signature of the QCD phase transition and the QCD critical point at high baryon density region are the main goals of the Beam Energy Scan (BES) programs at Relativistic Heavy-Ion Collider (RHIC)~\cite{bes}  and the Super-Proton Synchrotron (SPS) facilities~\cite{NA61_SPS}. Recent years, the STAR and PHENIX collider experiments already took the data of Au+Au collisions at \sNN=7.7, 11.5 (not taken by PHENIX), 14.5, 19.6, 27, 39, 62.4 and 200 GeV and the NA61/SHINE fixed target experiment has taken the data of p+p, Be+Be and Ar+Sc collisions at \sNN=5, 6.2, 7.6, 8.7, 12.3 and 17.3 GeV.   
\begin{figure}
\begin{center}
\includegraphics*[scale=0.7]{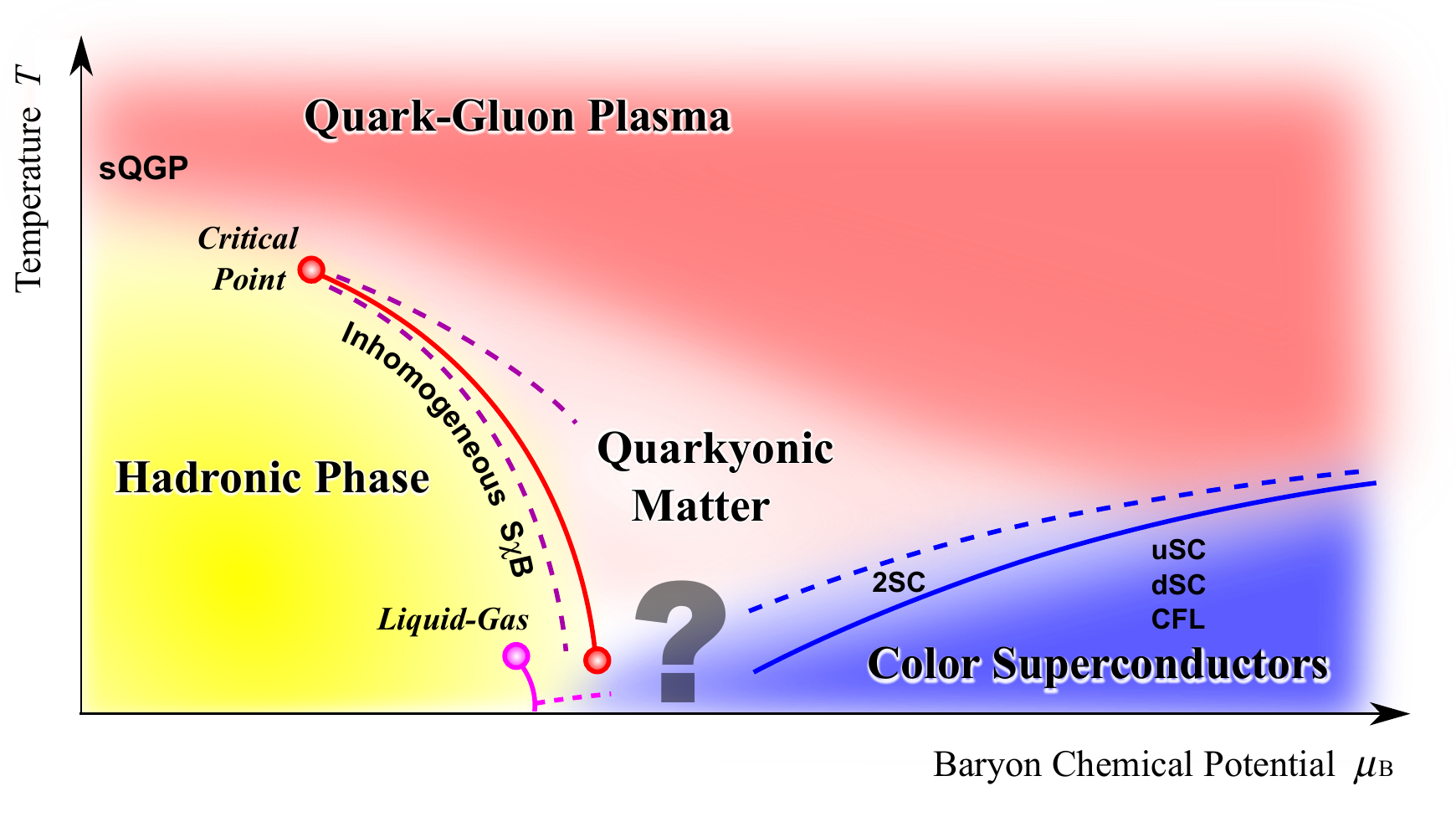}\\
\vspace{-0.5cm}
\caption{ (Color Online) The  QCD phase diagram~\cite{2011_Fukushima_phase}.}
\label{fig:phase}
\end{center}
\end{figure}

Experimentally, we aimed to explore the QCD phase structure at high baryon density regions and try to answer the following three key questions with the BES programs:
1). Can we see the turning off the promising QGP signatures established at top RHIC energy ? 2). Can we observe the first order phase transition at finite baryon density ? 
3). Can we discover the QCD critical point ?

\section{Results}
In the following sections, some experimental highlights from the beam energy scan will be presented.
At the beginning, we show the energy dependence of the freeze-out conditions and transverse dynamics. Measurements of nuclear modification factor $R_{CP}$ will be discussed in the section~\ref{sec:Rcp}, which is widely used as an observable for QGP formation at the RHIC top energy. Section~\ref{sec:v1} shows the results from directed flow and HBT radii measurements.  The energy dependence of the third order flow harmonics of charged particles will be discussed in the section~\ref{sec:v3}, which is sensitive to the lifetime of QGP and the pressure gradient in the system. Section~\ref{sec:moments} discussed the higher order event-by-event fluctuations of conserved quantities, respectively, which are used to search for the QCD critical point. Finally, we discuss the STAR detector upgrades and the prospective for the future beam energy scan programs. 

\subsection{Freeze-out conditions and transverse dynamics}  \label{sec:ch_freeze}
Freeze out dynamics are important bulk properties of the hot and dense nuclear matter created in heavy-ion collisions~\cite{1996_PBM_Thermal,2006_Andronic_Hadron}. 
It maps out the regions we can access in the QCD phase digram, which provide a baseline for finding the QCD phase transition and critical point. The concept of chemical and kinetic freeze-out are corresponding to the stages when the inelastic and elastic interactions of hadrons cease, respectively. 
Once the system reaches chemical freeze-out and kinetic freeze-out after hadronization, particle yields and transverse momentum spectra ($p_{T}$) get fixed, respectively. 
The chemical freeze-out temperature ($T_{ch}$) is obtained by fitting the particle yield 
or particle ratios with thermal model~\cite{Bedanga_freezeout}, whereas the kinetic freeze-out temperature ($T_{kin}$) is obtained from the blast-wave model fitting of the particle $p_{T}$ spectra~\cite{1993_Schnedermann_Thermal}.  
\begin{figure}[htb]
\hspace{0.2cm}
\begin{minipage}[c]{0.5\linewidth}
\centering 
    \includegraphics[scale=0.35]{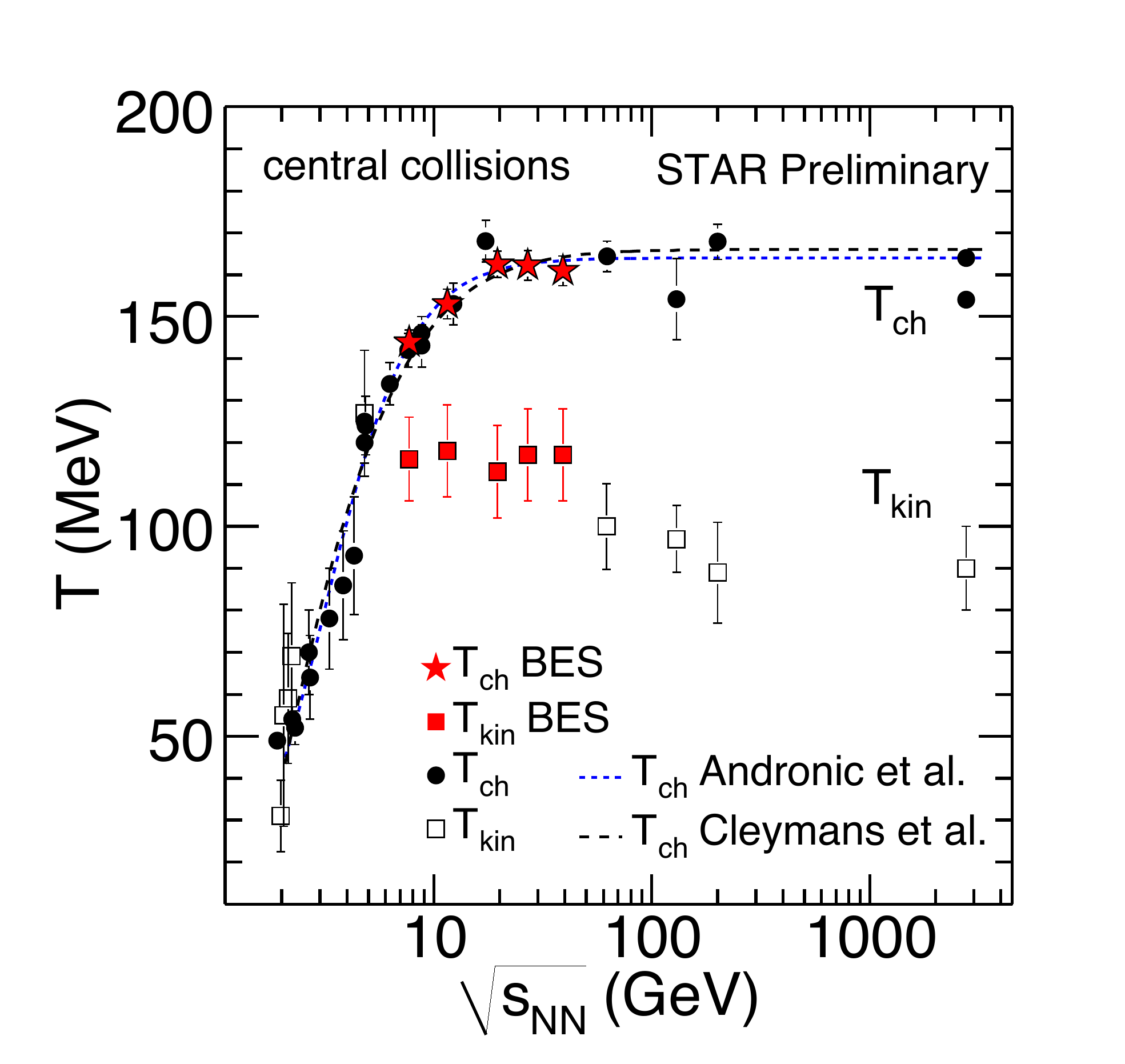}
    \end{minipage}
 \hspace{-0.5in}
  \begin{minipage}[c]{0.5\linewidth}
  \centering 
   \includegraphics[scale=0.4]{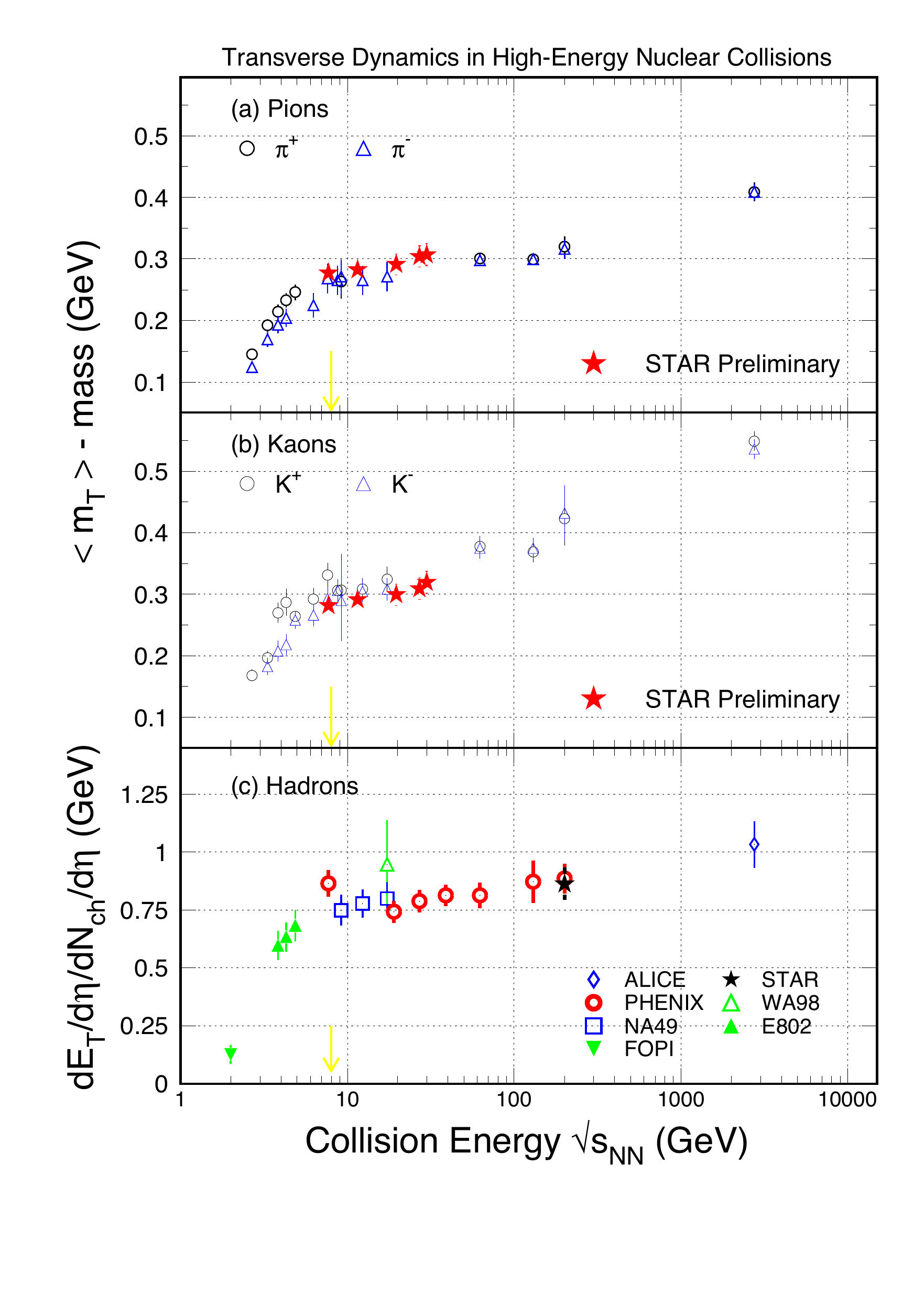}
   \vspace{1cm}
      \end{minipage}    
      \vspace{-1.8cm}
      \caption{ (Color online) Left: Energy dependence of 
chemical and kinetic freeze-out temperatures for central nuclear collisions. The dashed curves represent various
theoretical predictions.  Right: Energy dependence of the $<m_{T}>-m$ for pion and kaon, and transverse energy per charge multiplicity for different experiments.}  \label{fig:freeze}
\end{figure}

Figure \ref{fig:freeze} left shows the energy dependence of the chemical and kinetic freeze-out temperature $T_{ch}$ and $T_{kin}$ for central heavy-ion collisions from different experiments~\cite{2014_Lokesh_Systematics}. The chemical freeze-out temperature monotonically increases with {\sNN}  and then saturates at about \sNN=10 GeV, with a limiting temperature $T_{limt}\sim$160 MeV. The values of $T_{ch}$ and $T_{kin}$  are very close at energies below \sNN=7 GeV, whereas the $T_{kin}$ starts to deviate below the $T_{ch}$ with {\sNN} $\ge$ 7 GeV. The separation between chemical and kinetic freeze-out temperature becomes larger when going towards higher energies, which may be due to the stronger collective flow at higher energies. 
The $T_{kin}$ is nearly constant around the energies \sNN=$7.7-39$ GeV and slightly decrease towards the LHC energies. Fig.\ref{fig:freeze} right shows the $<m_{T}>-m$ as a function of \sNN\ for $\pi^{\pm}$, $K^{\pm}$~\cite{2014_Lokesh_Systematics} and the energy dependence of transverse energy per charged hadron multiplicity from different experiments~\cite{2014_Soltz_PHENIX}. These observables are also related to the transverse expansion dynamics of the system. 
The $<m_{T}>-m$ for $\pi^{\pm}$, $K^{\pm}$ increase rapidly at lower energies and almost remains constant around \sNN=7.7-39 GeV,  then increase again up to LHC energies. The similar behavior is observed for the transverse energy per charged hadron multiplicity. If the system was in thermal equilibrium, the quantity  $<m_{T}>-m$ is related to the temperature of the system and the energy \sNN\ can be related to the entropy of the system as $dN/dy \propto log($ \sNN $)$. Thus, the constant behavior of $<m_{T}>-m$  as a function of \sNN\ could be related to the first order phase transition~\cite{1982_Hove_FlatmT}.

\subsection{Nuclear modification factor R$_{CP}$} \label{sec:Rcp}
The nuclear modification factor $R_{CP}$ is widely used to quantify the suppression of high transverse momentum ($p_{T}$) hadrons, which 
provides the evidence for the energy loss of the energetic partons in hot and dense medium created in heavy-ion collisions, namely, jet quenching.  
This is one of the key signatures for the formation of QGP at the top RHIC energy~\cite{2005_STARwhitepaper}. It is defined as the ratio of yields in central collisions to yields in peripheral collisions and both yields are scaled by corresponding number of binary nucleon-nucleon collisions $N_{coll}$. 
\begin{figure}[htb]
\hspace{1.2cm}
\begin{minipage}[c]{0.5\linewidth}
\centering 
    \includegraphics[scale=0.4]{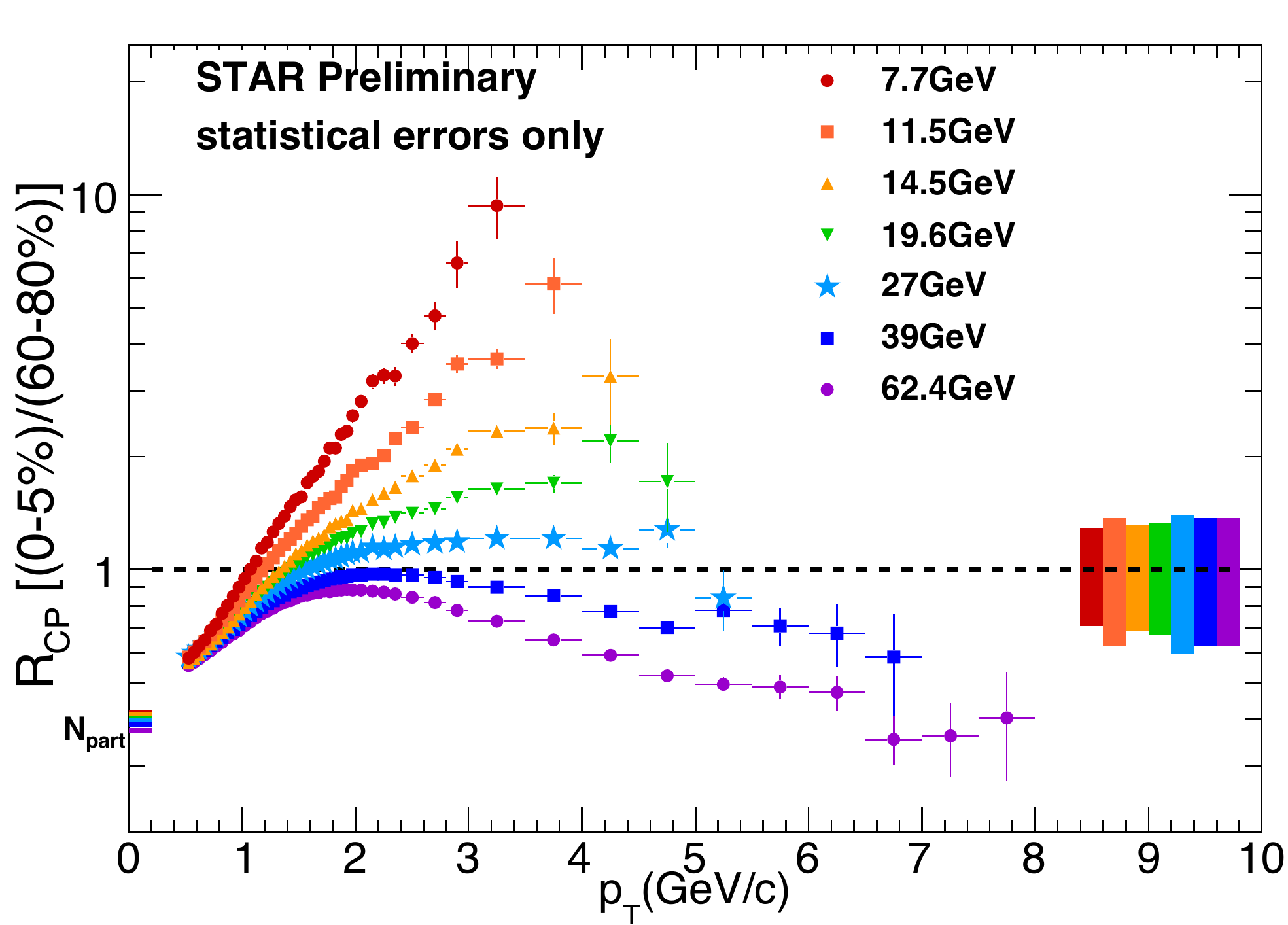}
    \end{minipage}
  \hspace{-0.5in}
  \begin{minipage}[c]{0.5\linewidth}
  \centering 
   \includegraphics[scale=0.35]{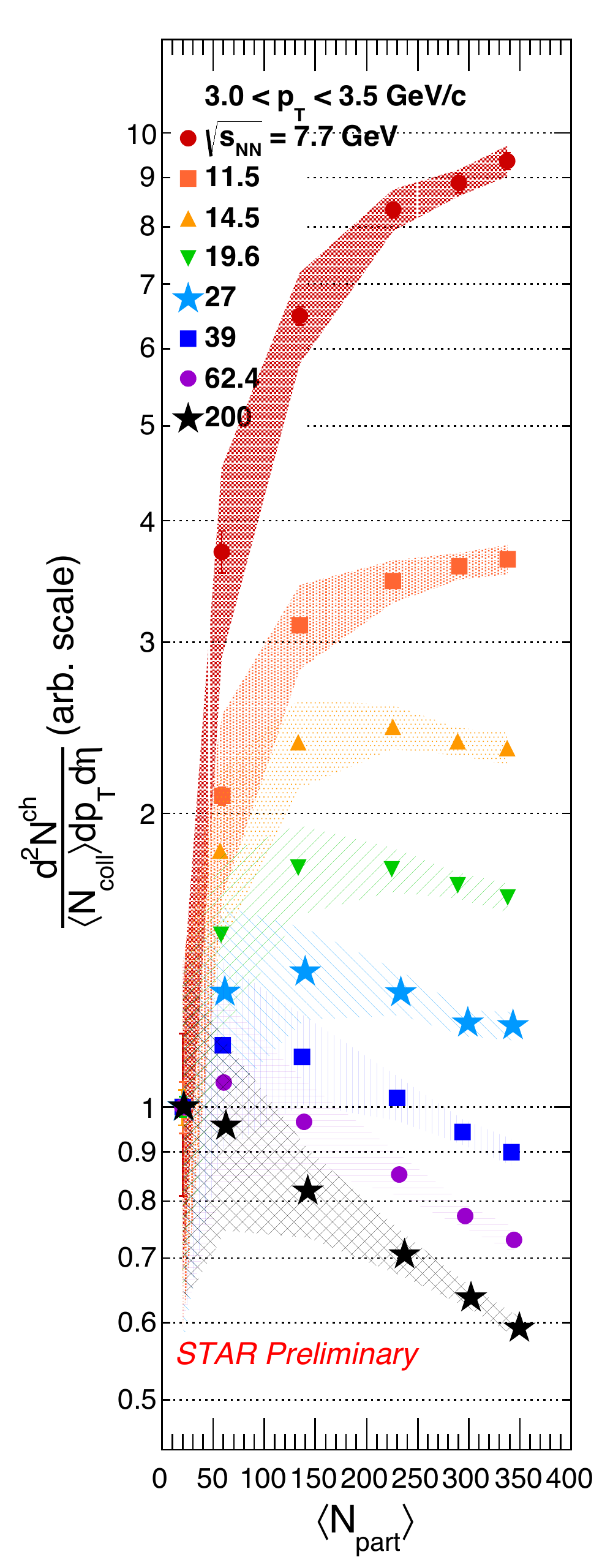}
      \end{minipage}    
      \vspace{-0.4cm}
      \caption{ (Color online) Left: The nuclear modification factor $R_{CP}$ as a function of the transverse momentum \pt\ for charged hadrons from RHIC BES data measured by STAR. 
       Right: the number of binary collisions scaled yield for charged hadrons as a function of the average number of participant nucleons $<N_{part}>$ for the \pt\ range $3.0<{\pt}<3.5$ GeV/c from the RHIC BES energies. } \label{fig:Rcp}
      \label{fig:Rcp}
\end{figure}
The $R_{CP}$ of charged hadrons are expected to be less than unity at high \pt\ due to parton energy loss in the dense medium and can serve as a turn off signature for the QGP formation at low energies and/or peripheral collisions.

Figure \ref{fig:Rcp} left shows the $R_{CP}$ of charged hadrons as a function of \pt\ for 0-5\%\ most central Au+Au collision at \sNN=7.7-200 GeV~\cite{2015_Horvat_QM}. Obviously, at high \pt\ (\pt $>$3 GeV/c), there shows a clear suppression for 39, 62.4 and 200 GeV, whereas 
the $R_{CP}$ is larger than unity for other low energies. However, this observation can not lead to the conclusion that there has no suppression effects at low energies as the initial state effects, such as cronin effect, 
may begin to dominat at lower energies. The enhancement of $R_{CP}$ caused by cronin effects will certainly compete with the suppression effects and make it more complicated to quantify the parton energy loss. A more effective way to 
demonstrate the interplay between enhancement and suppression effects is to show the centrality dependence of $R_{CP}$ of charged hadrons for high \pt\ range, i.e, $3<\pt<3.5 $ GeV/c as shown in Fig.\ref{fig:Rcp} right~\cite{2015_Horvat_QM}.  It shows that 
the suppression effects become more dominate in the central Au+Au collisions down to \sNN=14.5 GeV, whereas the cronin effects dominate at 7.7 and 11.5 GeV. Those may indicate that the lifetime and/or energy density of QGP formed in the central collisions at lower energies 
are shorter and smaller than high energies.  To obtain more clear understanding of the physics of $R_{CP}$, we still need to have more detailed theoretical comparison with data and more statistics to study the higher \pt\ above 4 GeV/c at low energies.

\subsection{Directed flow ($v_{1}$) and HBT Radii} \label{sec:v1}
\begin{figure}[htb]
\hspace{0.3cm}
\begin{minipage}[t]{0.5\linewidth}
\centering 
    \includegraphics[scale=0.4]{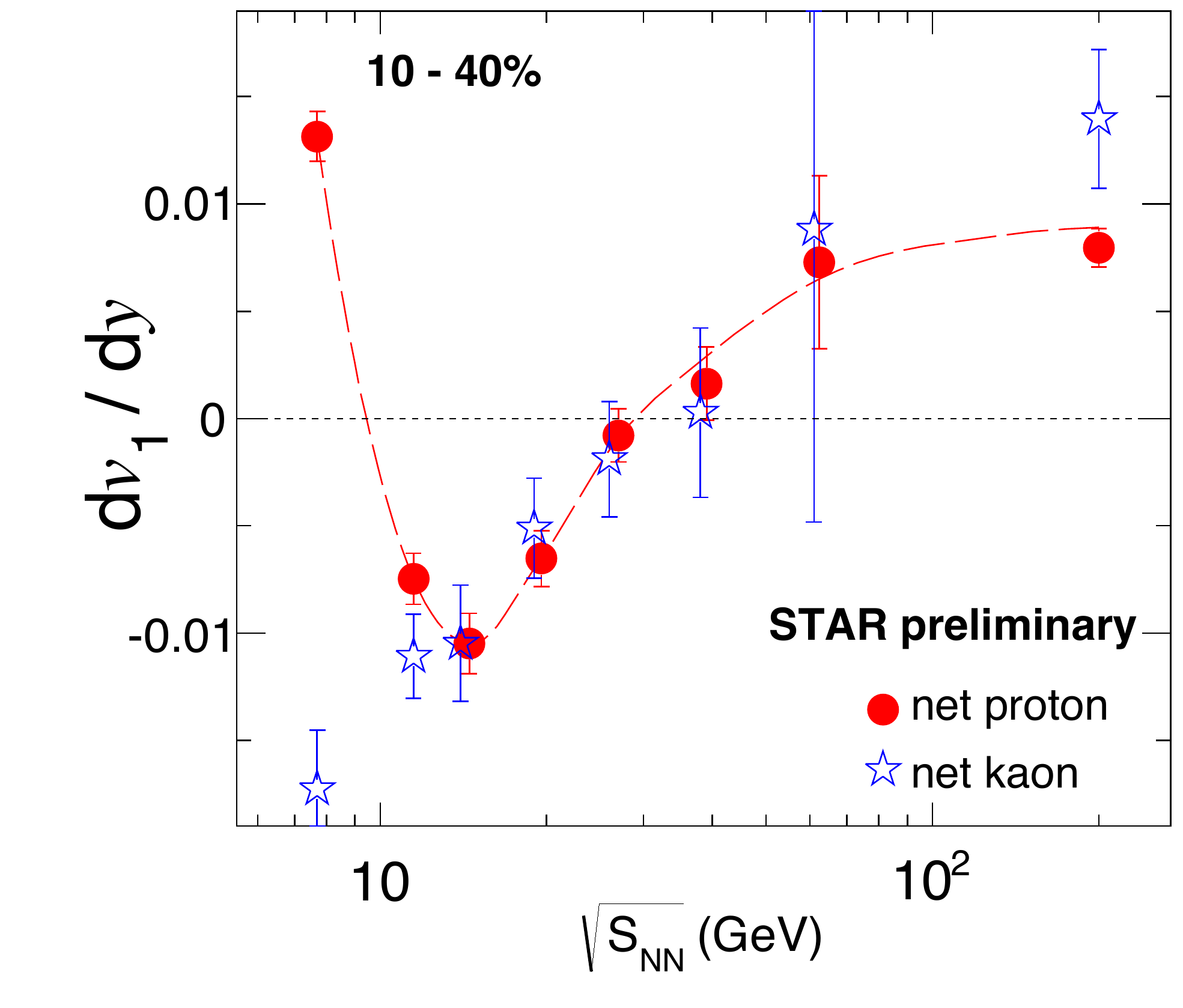}
    \end{minipage}
 \hspace{-0.3in}
  \begin{minipage}[t]{0.5\linewidth}
  \centering 
   \includegraphics[scale=0.6]{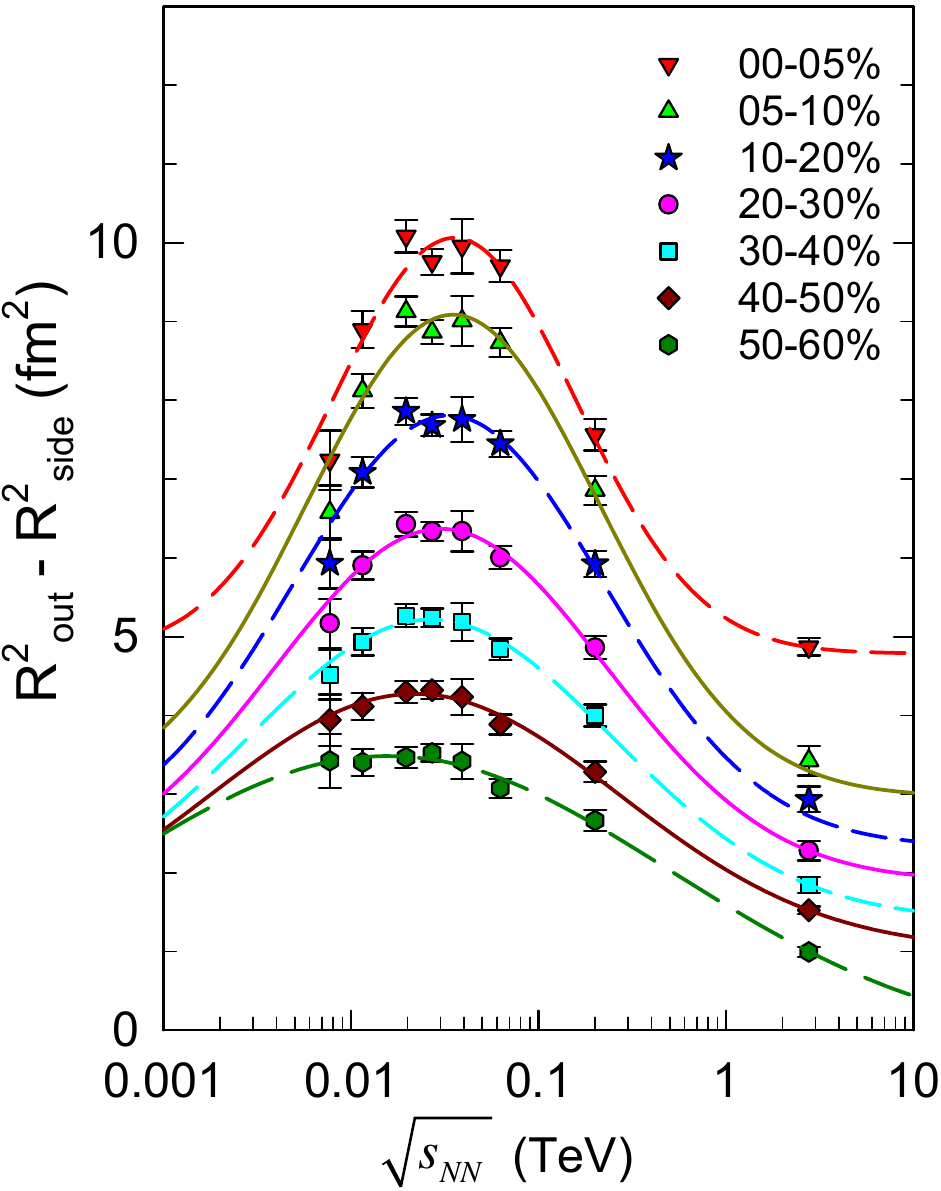}
   \vspace{1cm}
      \end{minipage}    
      \vspace{-1cm}
      \caption{ (Color online) Left: Energy dependence of directed flow slope for net-proton and net-kaon from RHIC BES energies measured by STAR. 
            Right: Energy dependence of HBT radii $R^{2}_\mathrm{out}-R^{2}_\mathrm{side}$ of charged pion from different experiments.  } \label{fig:v1}
\end{figure}

The energy dependence of directed flow of net-protons and/or protons have long been predicted as a sensitive probe for the first order phase transition in heavy-ion collisions~\cite{2004_Stocker_Collective}. Figure \ref{fig:v1} left shows slope of net-proton and net-kaon directed flow ($v_{1}$) as a function of \sNN\ measured by the STAR. The non-monotonic behavior is observed for the slope of the net-proton directed flow and the minimum is around \sNN=14.5 GeV~\cite{2015_Shanmuganathan_V1}. 
The slope of net-kaon directed flow shows a monotonic decrease trend with decreasing energy, which is consistent with the results of net-proton from \sNN=200 GeV down to 14.5 GeV and started to deviate at 11.5 GeV, significant difference is observed at 7.7 GeV.  The non-monotonic variation for slope of net-proton directed flow with \sNN\ could be related to the softening of equation-of-state due to the first order phase transition. However, final-state interaction effects, such as annihilation and hadronic potentials, may play important role in the observed final state directed flow. More detailed model comparisons and theoretical calculations are needed. Energy dependence of HBT radii $R^{2}_\mathrm{out}-R^{2}_\mathrm{side}$ of charged pion obtained from the STAR and PHNIEX experiment are shown in the right side of Fig.\ref{fig:v1}. 
It is predicted that the $R^{2}_\mathrm{out}-R^{2}_\mathrm{side}$  is proportional to the particle emission duration of the system. Intuitively, one might think the maximum value of the $R^{2}_\mathrm{out}-R^{2}_\mathrm{side}$  could be related to the presence of long lifetime fireball due to first order phase transition. Actually, the extracted exponents from the standard finite size scaling analysis suggests it is a second order phase transition and with the same universality class Z(2) in the 3D Ising model, which is consistent with a QCD critical point~\cite{2014_Roy_CP}. However, this needs more careful investigations before final conclusion.

\subsection{Third flow harmonics  ($v_{3}^{2}\{2\}$)} \label{sec:v3}
The $v_{3}^{2}\{2\}$  is the third harmonics in the fourier decomposition of the two-particle azimuthal correlations
, which is defined as $v_{3}^{2}\{2\}$=$cos<3\Delta \phi>$ and the $\Delta \phi =\phi_1-\phi_2$ is the relative azimuthal angle between two particles. Initially, the $v_{3}^{2}\{2\}$ was found to  originate from the initial geometry fluctuations.  In addition, recent hybrid model calculations show that the $v_{3}^{2}\{2\}$ is also very sensitive to a low viscosity QGP phase in the early stage~\cite{2013_Auvinen_v3}. This motivates us to use $v_{3}^{2}\{2\}$ to serve as a signature of formation of QGP.  Since the $v_{3}^{2}\{2\}$ is produced at early stage, it requires strong pressure gradient to propagate the initial fluctuation signal to the final state. Thus, the lifetime/viscosity of QGP as well as the strength of the pressure gradient in heavy-ion collisions dominate the size of the $v_{3}^{2}\{2\}$ signal. 
\begin{figure}[htb]
\hspace{0.2cm}
\begin{minipage}[t]{0.5\linewidth}
\centering 
    \includegraphics[scale=0.37]{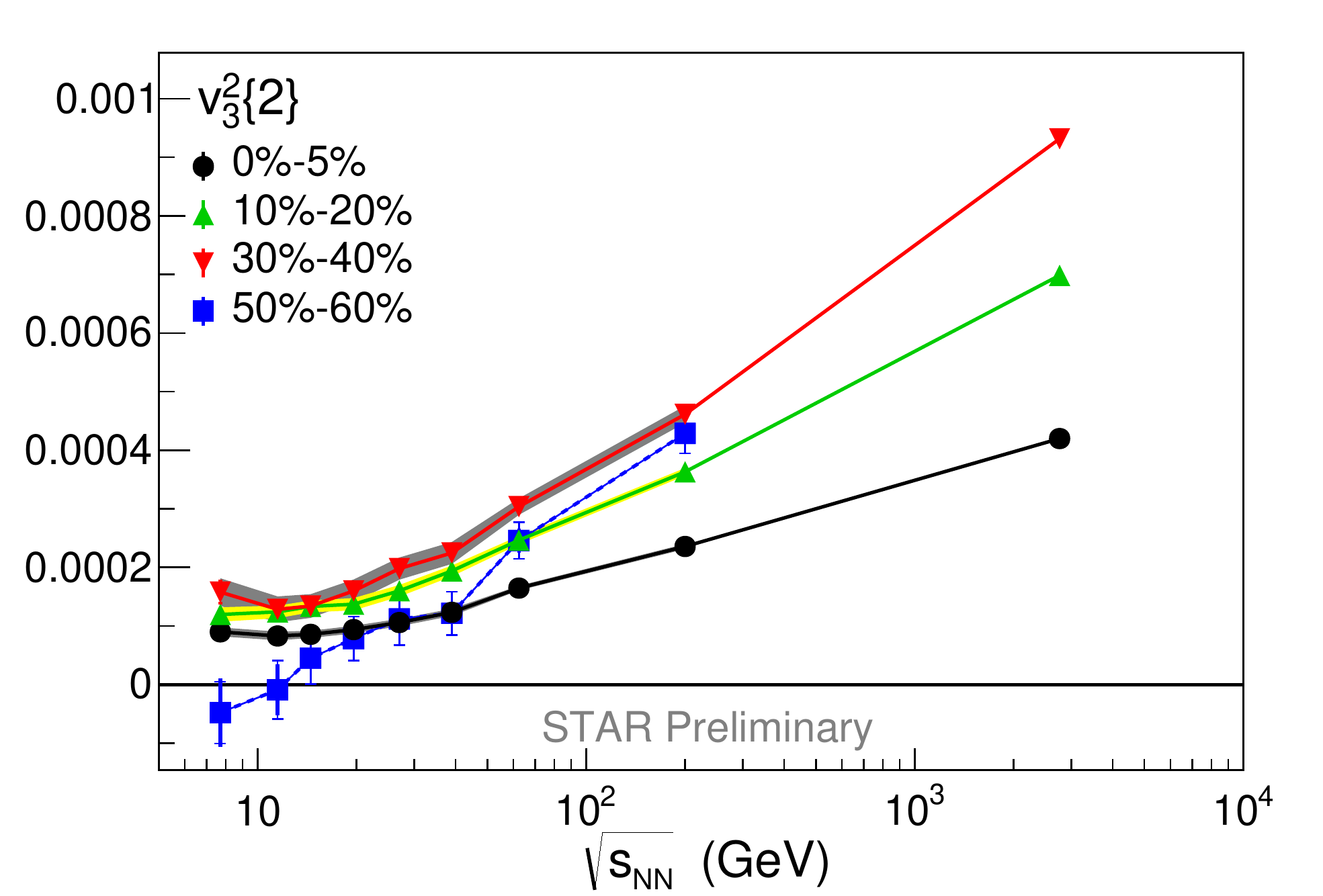}
    \end{minipage}
 \hspace{-0.5in}
  \begin{minipage}[t]{0.5\linewidth}
  \centering 
   \includegraphics[scale=0.32]{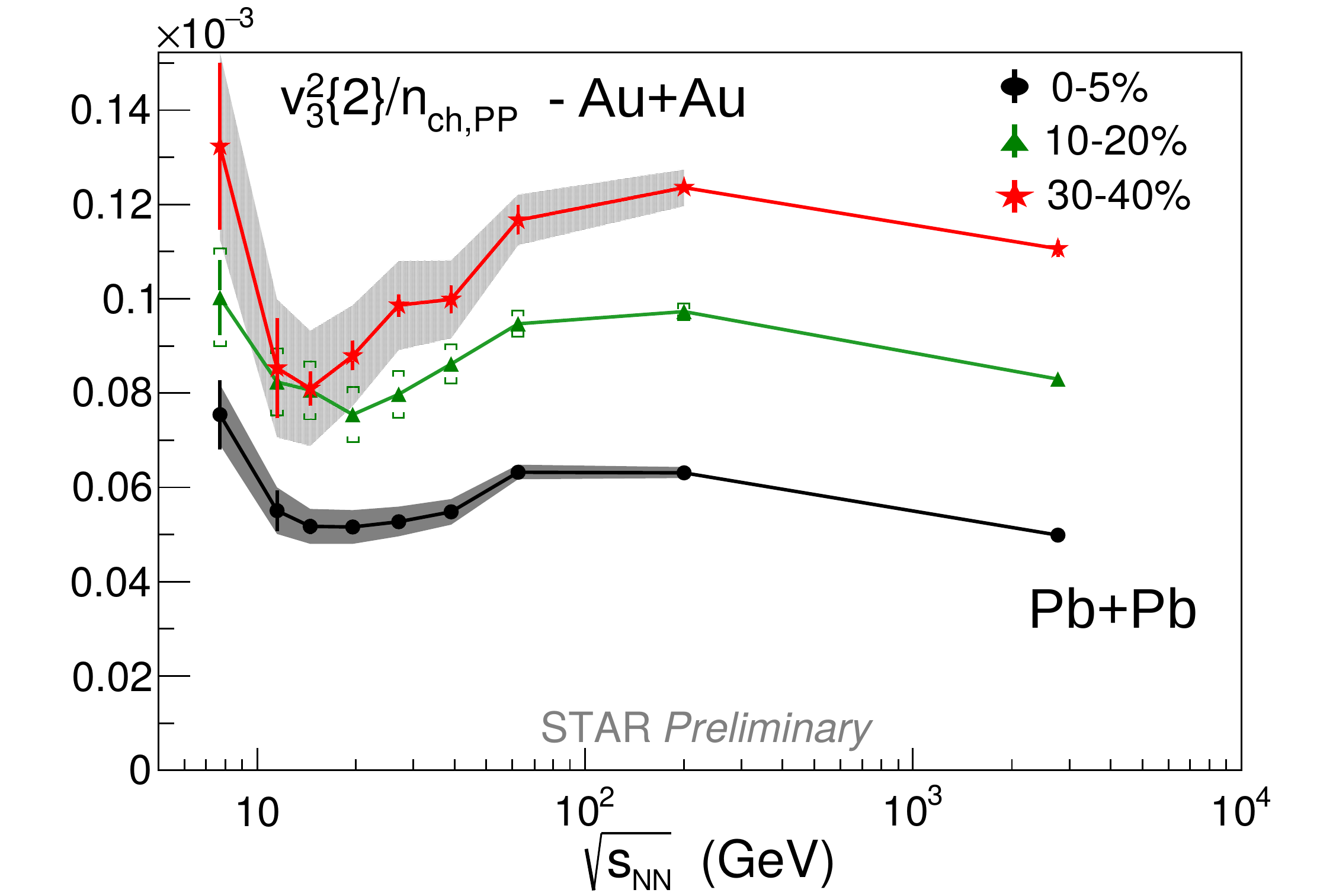}
      \end{minipage}    
      \vspace{-0.2cm}
      \caption{(Color online) Left: Energy dependence of $v_{3}^{2}\{2\}$ for four collision centralities from RHIC BES Au+Au collisions measured by STAR and the 
      2.76 TeV data points are from the Pb+Pb collisions measured by ALICE. Right: The $v_{3}^{2}\{2\}$ divided by the charged particle multiplicity per participant pair as a function of \sNN\ for three centralities.  } \label{fig:v3}
\end{figure}

Figure \ref{fig:v3} left shows the $v_{3}^{2}\{2\}$ as a function of the \sNN\ for four centralities~\cite{2015_Liao_V3}. The highest energy $v_{3}^{2}\{2\}$ is measured in the Pb+Pb collisions at 2.76 TeV by ALICE~\cite{2014_ALICE_V3}  and the others are from Au+Au collisions from the RHIC BES  measured by STAR.  At low energies, the $v_{3}^{2}\{2\}$ of  7.7 and 11.5 GeV at 50-60\% centrality become consistent with zero, which might indicate the absence of a low viscosity QGP phase in the low energies and/or peripheral collisions.  For more central collisions,  the values of $v_{3}^{2}\{2\}$  are positive and change little from 19.6 GeV to 7.7 GeV.
For the energies above 19.6 GeV,  the values of $v_{3}^{2}\{2\}$ linearly increase with the $log($\sNN$)$ for all of the four centralities.
Figure \ref{fig:v3} right shows \sNN\ dependence of the $v_{3}^{2}\{2\}$ scaled by the charged particle multiplicity per participant pair ${n_{ch,\mathrm{PP}}} = \frac{2}{{{N_\mathrm{part}}}}d{N_{ch}}/d\eta$ for three centralities. Experimentally, the ${n_{ch,\mathrm{PP}}}$ has been measured and monotonically increase with \sNN~\cite{2011_PHOBOS_Multiplicity}, which can be related to the energy density of the system. The $v_{3}^{2}\{2\}$/${n_{ch,\mathrm{PP}}}$ shows a local minimum around 20 GeV, which is the consequence of a relatively flat trend for $v_{3}^{2}\{2\}$ and monotonically increasing trend for the ${n_{ch,\mathrm{PP}}}$  in the energy range 7.7 $<$\sNN $<$ 20 GeV. Physics wise, the $v_{3}^{2}\{2\}$/${n_{ch,\mathrm{PP}}}$ should reflect the ability of the system to convert the initial geometry fluctuations to the final state. Thus, the local minimum in $v_{3}^{2}\{2\}$/${n_{ch,\mathrm{PP}}}$ could indicate an anomalous low pressure inside the matter created in the collisions near \sNN=20 GeV, where a minimum is also observed for the slope of net-proton directed flow. Apparently, these observations can be interpreted by softening of equation-of-state due to presence of the first order phase transition. However, conclusions only can be made after carrying out careful theoretical and model studies for the dynamical evolution of the system including the physics of first order phase transition at finite $\mu_{B}$. 

\subsection{Net-proton number fluctuations}  \label{sec:moments}
Fluctuations of conserved quantities, such as baryon (B), charge (Q) and strangeness (S) numbers, have been proposed as a sensitive probe to search for the signature of the QCD critical point in heavy-ion collisions~\cite{Asakawa}. These fluctuations are sensitive to the correlation length ($\xi$)~\cite{Asakawa} and can be directly connected to the susceptibility of the system computed in theoretical calculations, such as Lattice QCD~\cite{freezeout,MCheng2009,science} and HRG models~\cite{HRG}. The STAR experiment has measured various order fluctuations of net-proton ($N_{p}-N_{\bar{p}}$, proxy for net-baryon), net-charge and net-kaon (proxy for net-strangeness) numbers in the Au+Au collisons at \sNN=7.7, 11.5, 14.5, 19.6, 27, 39, 62.4 and 200 GeV~\cite{2010_NetP_PRL, STAR_BES_PRL,2014_Luo_CPOD}. 
\begin{figure}[htb]
\hspace{-0.5cm}
\begin{minipage}[c]{0.35\linewidth}
\centering 
    \includegraphics[scale=0.25]{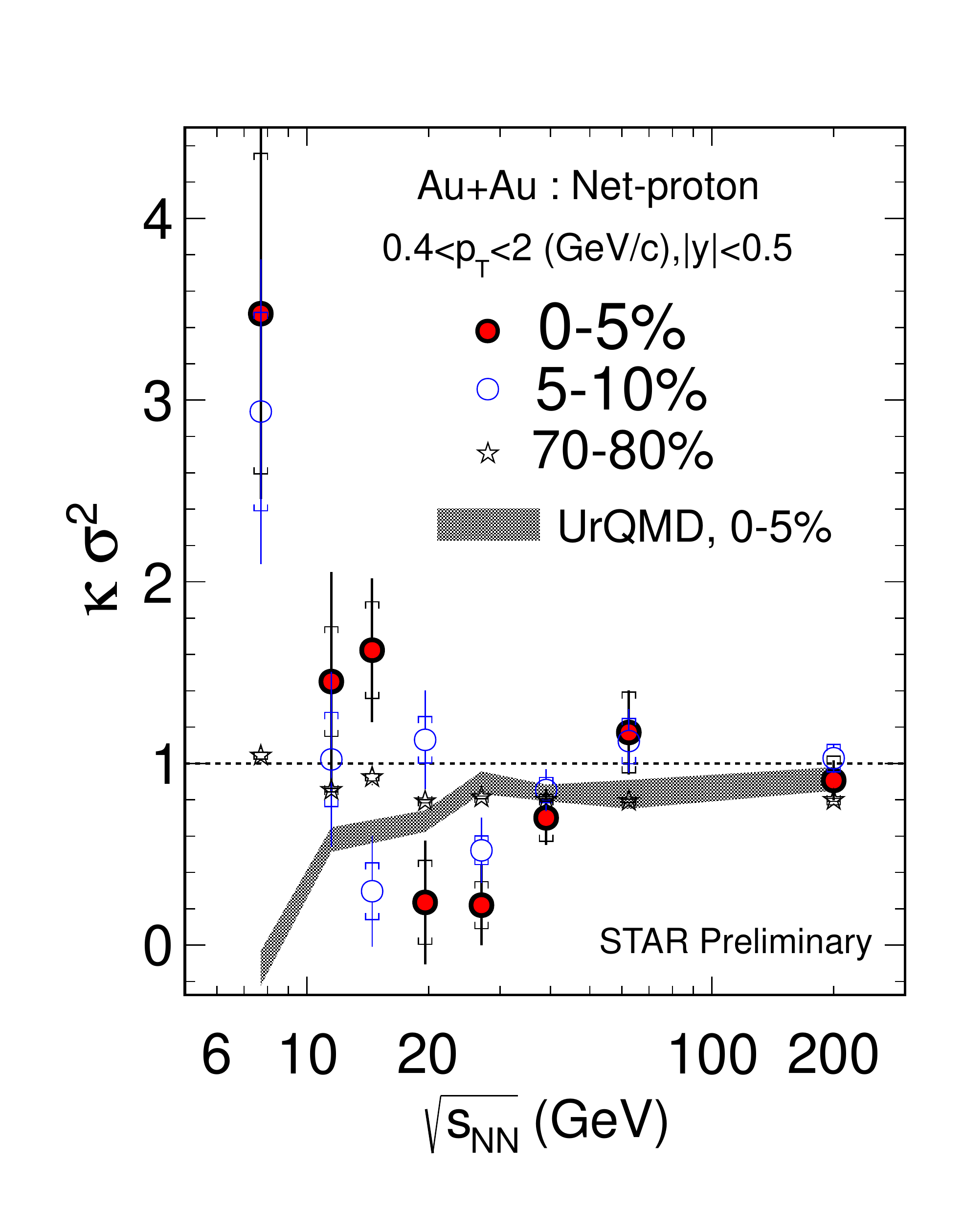}
    \end{minipage}
  \hspace{-0.35in}
  \begin{minipage}[c]{0.35\linewidth}
  \centering 
   \includegraphics[scale=0.27]{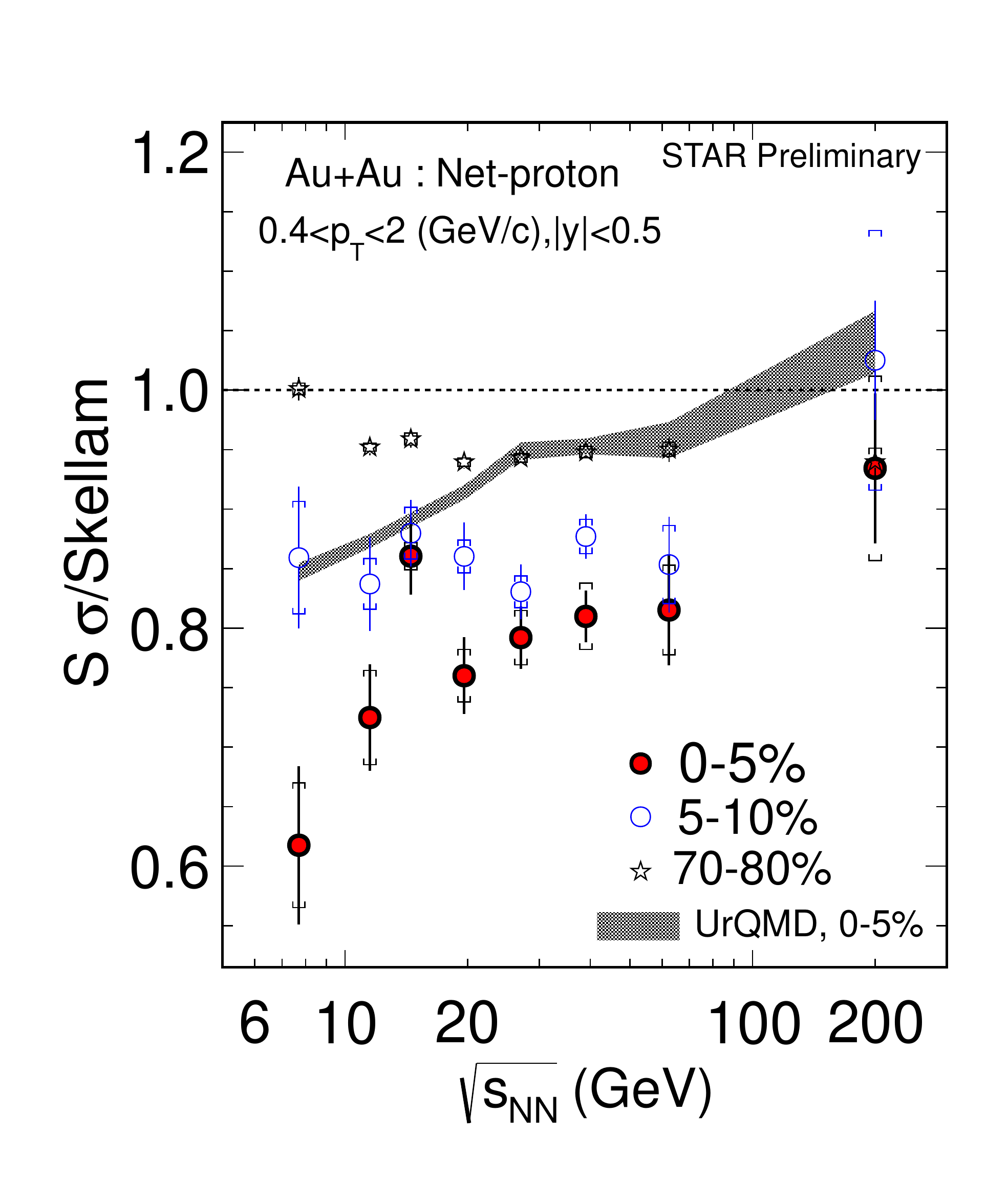}
      \end{minipage}    
       \hspace{-0.2in}
        \begin{minipage}[c]{0.35\linewidth}
  \centering 
   \includegraphics[scale=0.34]{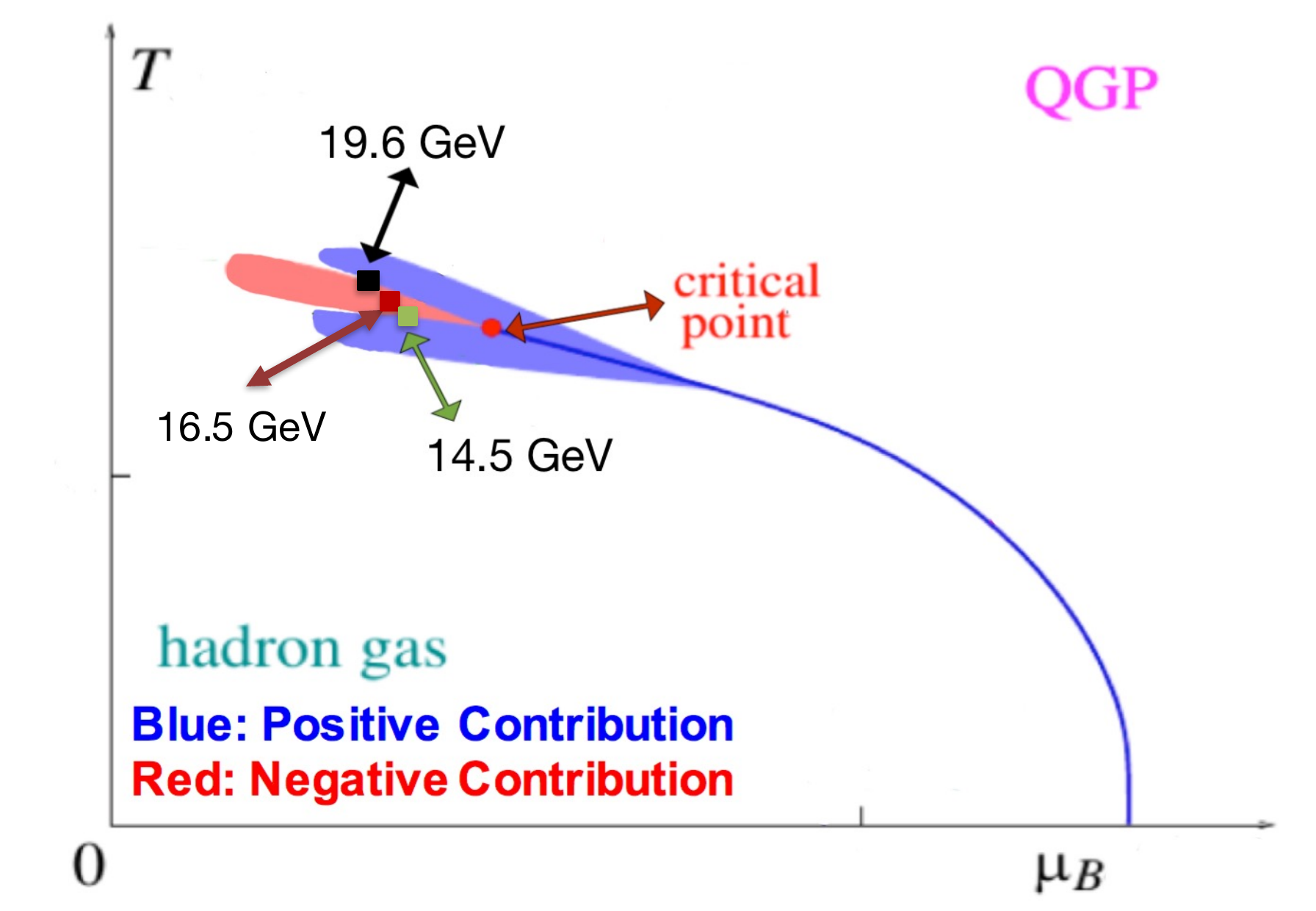}
      \end{minipage} 
      
      \vspace{-0.4cm}
      \caption{ (Color online) Left: Energy dependence of {\KV} of net-proton distributions and Middle: {\SD} divided by Skellam (Poisson) expecations for 0-5\%, 5-10\% and 70-80\% centralities of Au+Au collisions at \sNN=7.7, 11.5, 14.5, 19.6, 27, 39, 62.4, 200 GeV measured by STAR. The experimental data is compared with Poisson expectations (dashed lines) and the UrQMD transport model calculations (shade bands ). The statistic and systematic errors are plotted as vertical bar and brackets, respectively. Right:  A schematic sketch for theoretically predicted negative(red)/positive(blue) critical contribution regions for {\KV} near the QCD critical point and possible chemical freeze-out regions for Au+Au collisions 14.5 (green), 16.5 (purple) and 19.6 GeV (black).} \label{fig:cumulants_energy}
\end{figure}

Figure  \ref{fig:cumulants_energy} left shows the efficiency corrected \KV\ of net-proton distributions as a function of \sNN\ for 0-5\%, 5-10\% and 70-80\% centralities of Au+Au collisions measured by STAR~\cite{2014_Luo_CPOD,2015_Jochen_QM}. The protons and anti-protons numbers are measured with transverse momentum $0.4<\pt<2$ GeV/c and at mid-rapidity $|y|<0.5$. The \KV\ shows a clear non-monotonic variation with \sNN\ for 0-5\% centrality with a minimum around 20 GeV.  Above 39 GeV, the values of \KV\ are close to the unity for both central and peripheral collisions and deviate significantly below unity for the 0-5\% most central collisions at 19.6 and 27 GeV, then become above unity at 0-5\% centrality in the energies below 19.6 GeV.  Another intriguing structure observed in \sNN\ dependence for the \KV\ of net-proton distributions in Au+Au collisons is the so called "Oscillation".  Namely, the oscillation is a structure that represents two observations, the so called "separation" and "flipping".  The "separation" means that at 14.5 and 19.6 GeV, the absolute difference of {\KV} values between the 0-5\% and 5-10\% centralities are large comparing to other energies. The "flipping" means that the values of the {\KV} at 0-5\% and 5-10\% with respect to unity shows a flipping at 14.5 and 19.6 GeV, i.e., for 14.5 GeV, the value of \KV\ at 0-5\% is larger than unity while \KV\ of 5-10\% centrality is smaller than unity, but it is inverse at 19.6 GeV. Fig. \ref{fig:cumulants_energy} middle shows the \SD\ of net-proton distributions divided by the skellam baselines~\cite{QM2014_baseline} as a function of \sNN\ for three centralities measured by STAR. For 0-5\% centrality,  the \SD/Skellam shows significantly deviate below unity and monotonically decrease with decreasing energies except for showing a peak at 14.5 GeV. The results from UrQMD model without critical point, show a monotonic decreasing trend as a function of \sNN\ and can not reproduce the structures observed in \KV\ and \SD/Skellam of net-proton distributions measured by STAR.

Theoretically, there is a critical region around the critical point and its size depends on how the correlation length ($\xi$) of the system varies with the distance from the critical point ($\xi(d)$). When the thermodynamic conditions ($T$ and $\mu_{B}$) of hot and dense nuclear matter created in heavy-ion collisions enter into the critical region and get closer to the critical point, experimental probes will "feel" about the critical fluctuations and receive critical contributions. For different observables, the behavior near the QCD critical point could be very different.  As illustrated in Fig.\ref{fig:cumulants_energy} right, according to the theoretical calculations~\cite{Neg_Kurtosis,2014_Bengt_flu,2015_JianDeng_fluctuation,2016_fluctuations_Huichao}, the critical contributions to the observable {\KV} consist of positive and negative two sub-regions displayed as blue and red in Fig.\ref{fig:cumulants_energy} right, respectively.  Experimentally, in addition to the statistical baseline unity, the sign of critical contributions for the {\KV} depends on the relative position of chemical freeze-out and the critical contribution region. The observed oscillation structure in \KV\ of net-proton distribution of 0-5\% and 5-10\% centralities at 14.5 and 19.6 GeV may indicate that the corresponding chemical freeze-out  $T$ and $\mu_{B}$ of these two energies have already entered into the critical region and received positive and/or negative contributions from critical fluctuations due to near the QCD critical point. Although, this is only qualitative arguments, more quantitative predictions can be made in the future by careful modeling for the dynamical evolution of heavy-ion collisions and the shape/size of the critical region. Furthermore, current statistical error bars are still large and the confirmation of the "oscillation" structure requires more statistics, which can be realized in the second phase of the beam energy scan program at RHIC.

Finally, regarding the intriguing structures observed in the \sNN\ dependence of \KV\ of net-proton distributions measured by the STAR experiment, I have the following comments, suggestions and predictions:
1). In heavy-ion collisions, event-by-event fluctuations of chemical freeze-out  $T$ and $\mu_{B}$  would be important in explaining the very different behavior of \KV\ observed in 0-5\% and 5-10\% centralities, and should be considered into dynamical modelling of the critical signals in heavy-ion collisions. 2). Besides having more statistics at 14.5 and 19.6 GeV in RHIC BES-II,  I would strongly suggest to make a finer energy scan between 14.5 and 19.6 GeV to verify and map out the details of the possible critical region near the QCD critical point.  3). Specifically,  one can take the energy point \sNN=16.5 GeV, which is with the similar chemical freeze-out $\mu_{B}$ interval from the 14.5 and 19.6 GeV. For 0-5\% collisions at 16.5 GeV, the {\KV} of net-proton distributions could have larger deviation from unity than 19.6 GeV due to closer to critical point. At 16.5 GeV, the {\KV} of 5-10\% centrality would have no big deviation from the 0-5\% centrality unlike at 14.5 and 19.6 GeV, since the chemical freeze-out points at 16.5 GeV might entirely fall into the negative contribution region (red) as shown in Fig.\ref{fig:cumulants_energy} right. 4). To check whether the large increasing for {\KV} at 7.7 and 11.5 GeV is due to large positive critical contribution or not, we could have at least one data point at even lower energies, i.e \sNN$< $8 GeV, to look for falling of the ${\KV}$ value below unity due to running outside of the critical region.

\subsection{Detector upgrades and beam energy scan phase II at RHIC}  \label{sec:besII}
Since many intriguing structures have been observed for the energy dependence of various observables in the first phase of RHIC beam energy scan (BES-I) programs, RHIC has planned to conduct second phase of the beam energy scan (BES-II) in the years 2019-2020 and focusing on energies below \sNN=20 GeV~\cite{BESII_WhitePaper}. It is to accumulate more statistics to further confirm the interesting structures or trends observed in the BES-I. This will provide us opportunity to draw a solid conclusion and have more complete/consistent physics picture from various experimental measurements with high precision. Thus, upgrades for detector and accelerator have been launched at RHIC.
\begin{figure}[htb]
\hspace{0.2cm}
\begin{minipage}[t]{0.5\linewidth}
\centering 
    \includegraphics[scale=0.5]{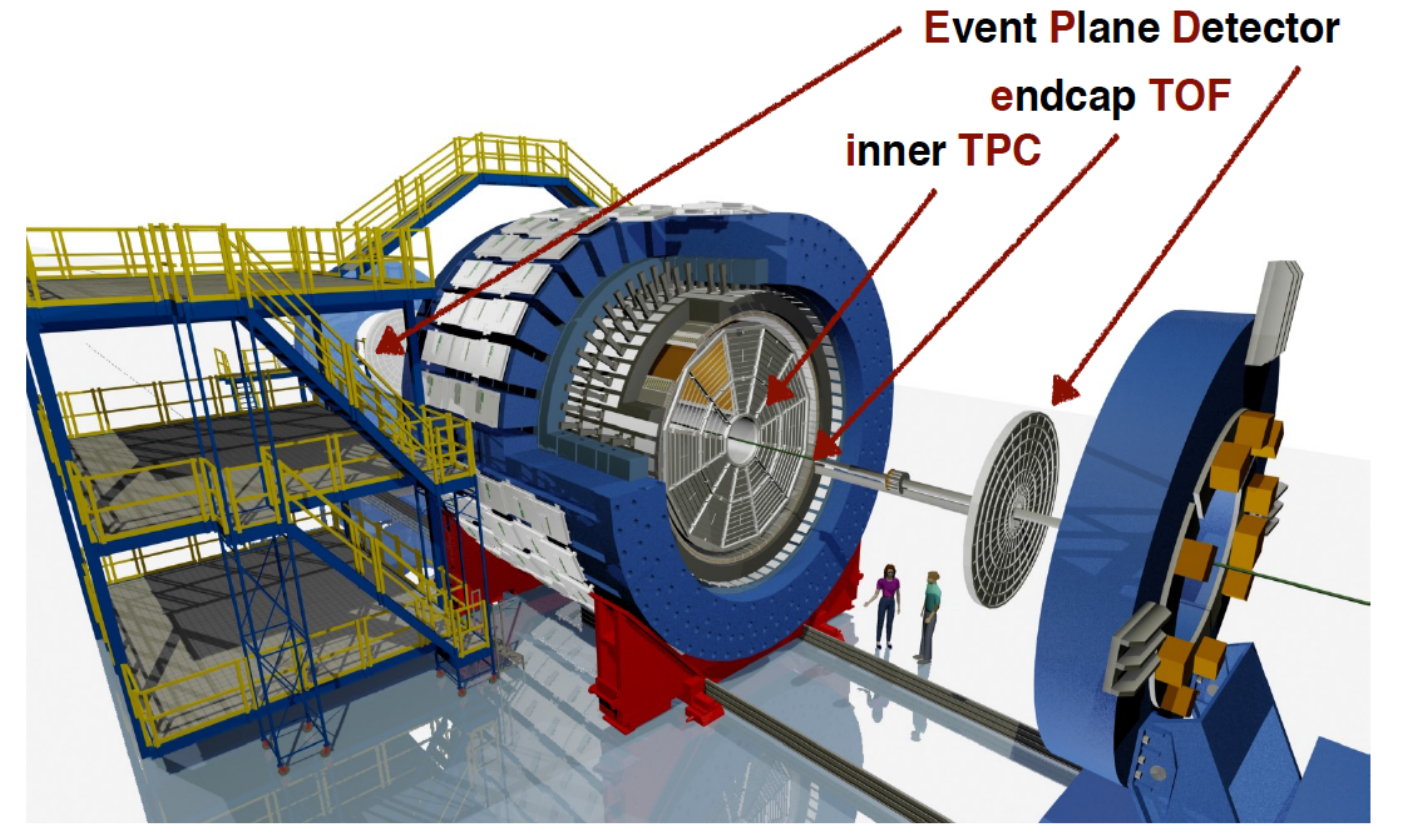}
    \end{minipage}
\hspace{-0.3in}
  \begin{minipage}[t]{0.5\linewidth}
  \centering 
   \includegraphics[scale=0.55]{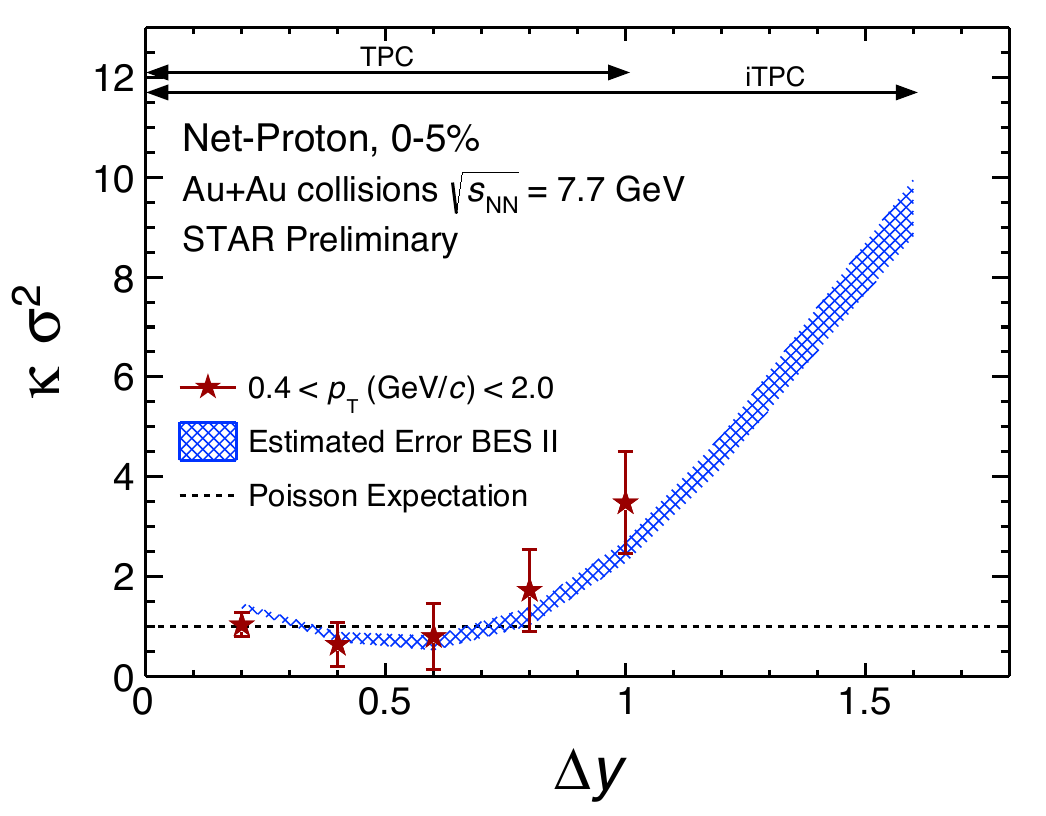}
      \end{minipage}    
      \vspace{-0.2cm}
      \caption{ (Color online) Left: Upgrades of the STAR detector for the second phase of beam energy scan at RHIC. Right: Rapidity coverage dependence of the {\KV} of net-proton distribution in 0-5\% central Au+Au collisions at \sNN=7.7 GeV. The blue band shows the expecting trend and statistical error for net-proton {\KV} at BES-II. For this analysis, the rapidity coverage can be extend to $|y|<0.8$ with iTPC upgrades. } \label{fig:BESII}
\end{figure}
The stochastic electron cooling technique and long beam bunches will be applied to accelerate gold beam, which will improve the luminosity by a factor about 5-15 compared to BES-I.  As the luminosity will decrease as decreasing the energy, the improvement of the luminosity is more important at low energies, such as 7.7 and 11.5 GeV.  A goal of about 100 million minimum bias events are expected to collect for the Au+Au collisions at \sNN=7.7 GeV by the STAR experiment, which is the lowest energy in collider mode of BES-II. To explore the QCD phase structure at even higher baryon density, running of STAR detector at the fixed target mode has been proposed. Test runs of fixed target mode were successfully conducted and preliminary results have been obtained for Au+Au collisions at \sNN=3.9 and 4.5 GeV collected in 2014 and 2015, respectively. In the future BES-II, fixed target mode collisions allow us to have energy coverage from \sNN=3 GeV ($\mu_{B}$=720 MeV) up to 7.7 GeV.

In Fig.\ref{fig:BESII} left, the inner TPC (iTPC) of STAR is being upgraded to improve the energy loss resolution and can extend the pseudo-rapidity acceptance from $|\eta|<1$ to $|\eta|<1.5$~\cite{iTPC_proposal}. It is also planed to install a Time-of-Flight (eTOF) detector at the west end cap of the STAR TPC to extend the PID capability at forward region. The iTPC upgrade is important to test the criticality and study dynamical evolution of the fluctuations by looking at the rapidity coverage dependence for the fluctuations of conserved quantities~\cite{2015_Kitazawa_Rapidity}.  In Fig.\ref{fig:BESII} right, the blue band is the extrapolating from current measurements by assuming critical contributions ($\KV{\propto}N^{3}$~\cite{2015_Acceptance_Misha}). In the forward and backward region of STAR detector, a new Event Plane Detector (EPD) will be built and used to replace the old Beam-Beam Counters (BBC) for independent centrality and event plane measurements, which can reduce the auto-correlations in the measurements at mid-rapidity. Due to the discovery potential at high baryon density,  future experimental facilities beyond current running experiments at RHIC and SPS, are planed to be built to study the physics at
low energies. The fixed target heavy-ion collisions experiments Compressed Baryonic Matter experiment (CBM) at Facility for Anti-proton and Ion Research (FAIR) at GSI, Germany will cover the energy range \sNN=2-8 GeV and the Japan Proton Accelerator Research Complex (J-PARC) will cover the energy range \sNN=2-6.2 GeV. The collider mode heavy-ion collision experiment Multi Purpose Detector (MPD) at Nucleon based Ion Collider fAcility (NICA) will cover the energy range \sNN=4-11 GeV at JINR, Russia.

\section{Summary and Outlook}
Beam energy scan programs in heavy-ion collisions have been carried out by RHIC and SPS with the main goals of finding the signature of QCD phase transition and QCD critical point at high baryon density region. During past few years, we have found many intriguing non-monotonic structures in the energy dependence of various observables in Au+Au collisions, such as dips in the slope of net-proton directed flow and $v_{3}^{2}\{2\}$/${n_{ch,\mathrm{PP}}}$, peak in the HBT radii measurement, and oscillations in {\KV} of the net-proton distributions. All of these structures are observed at the similar energy region $14<${\sNN}$<20$ GeV, which suggests some interesting could happen there. However, one should keep in mind that in this energy range the baryon density and baryon to meson ratio also change a lot, which makes it much complicated to attribute the observed structures to the QCD phase transition or QCD critical point. In the near future, it would be very helpful to explain the low energies data by comparing with the results from hydrodynamics and/or hybrid models including the realistic equation-of-state at finite baryon baryon density. Experimentally, the second phase of the beam energy scan at RHIC has been planed in 2019-2020 with upgraded detectors and increased luminosity to explore the phase structure focusing on energies below 20 GeV with high precision. 
\section*{Acknowledgement}
X. Luo is grateful for the constructive and helpful discussion with Prof. Masayuki Asakawa, Peter Braun-Munzinger, ShinIchi Esumi, Bengt Friman, Kenji Fukushima, 
Rajiv Gavai, Sourendu Gupta, Frithjof Karsch, Masakiyo Kitazawa, Volker Koch, Bill Llope, Roy Lacey, Bedangadas Mohanty, Marlene Nahrgang, Hans Georg Ritter, Krzysztof Redlich, Misha Stephanov, Johanna Stachel and Nu Xu. 
The work was supported in part by the MoST of China 973-Project No.2015CB856901 and NSFC under grant No. 11575069 and 11221504.  








\bibliography{QM15_XiaofengLuo}
\bibliographystyle{unsrt}

\end{document}